\documentclass[a4paper]{amsart}

\RequirePackage{amsmath}
\RequirePackage{bm}
\RequirePackage{amssymb}
\RequirePackage{upref}
\RequirePackage{amsthm}
\RequirePackage{enumerate}
\RequirePackage{pb-diagram}
\RequirePackage{amsfonts}
\RequirePackage[mathscr]{eucal}
\RequirePackage{verbatim}
\RequirePackage{xr}
\RequirePackage{graphicx}
\usepackage{calc}
\usepackage{xspace}
\RequirePackage{color}
\RequirePackage{ifthen}
\RequirePackage{fancybox}

\newcommand{\cf}{cf.\@\xspace}

\newcommand{\al}{\alpha}

\newcommand{\ga}{\gamma}
\newcommand{\de}{\delta }
\newcommand{\e}{\epsilon}

\newcommand{\f}{\varphi}
\newcommand{\h}{\eta}

\newcommand{\lam}{\lambda}
\newcommand{\m}{\mu}

\newcommand{\s}{\sigma}

\newcommand{\D}{\varDelta}
\newcommand{\F}{\varPhi}
\newcommand{\Lam}{\varLambda}
\newcommand{\Om}{\varOmega}

\newcommand{\so}{{\mc S_0}}

\newcommand{\msp[1]}[1]{\mspace{#1mu}}

\newcommand{\R}[1][n+1]{{\protect\mathbb R}^{#1}}

\newcommand{\Cc}{{\protect\mathbb C}}

\newcommand{\N}{{\protect\mathbb N}}

\newcommand{\eR}{\stackrel{\lower1ex \hbox{\rule{6.5pt}{0.5pt}}}{\msp[3]\R[]}}
\newcommand{\eN}{\stackrel{\lower1ex \hbox{\rule{6.5pt}{0.5pt}}}{\msp[1]\N}}
\newcommand{\eO}{\stackrel{\lower1ex \hbox{\rule{6pt}{0.5pt}}}{\msc O}}

\newcommand\im{\implies}
\newcommand\ra{\rightarrow}

\newcommand\hra{\hookrightarrow}

\newcommand{\un}{\infty}
\newcommand{\A}{\forall}

\newcommand{\uu}{\cup}
\newcommand{\ii}{\cap}
\newcommand{\uuu}{\bigcup}

\newcommand{\uud}{ \stackrel{\lower 1ex \hbox {.}}{\uu}}
\newcommand{\uuud}[1]{ \stackrel{\lower 1ex \hbox {.}}{\uuu_{#1}}}
\newcommand\su{\subset}
\newcommand\Su{\Subset}

\newcommand{\sminus}[1][28]{\raise 0.#1ex\hbox{$\scriptstyle\setminus$}}

\newcommand{\eqv}{\Longleftrightarrow}

\newcommand{\abs}[1]{\lvert#1\rvert}
\newcommand{\absb}[1]{\Bigl|#1\Bigr|}
\newcommand{\norm}[1]{\lVert#1\rVert}

\newcommand{\spd}[2]{\protect\langle #1,#2\protect\rangle}

\newcommand{\tit}{\textit}

\newcommand{\tup}{\textup}

\newcommand{\mc}{\protect\mathcal}
\newcommand{\msc}{\protect\mathscr}

\providecommand{\bysame}{\makebox[3em]{\hrulefill}\thinspace}

\newcommand{\bt}{\begin{thm}}
\newcommand{\bl}{\begin{lem}}
\newcommand{\bc}{\begin{cor}}
\newcommand{\bd}{\begin{definition}}
\newcommand{\bpp}{\begin{prop}}
\newcommand{\br}{\begin{rem}}
\newcommand{\bn}{\begin{note}}
\newcommand{\be}{\begin{ex}}
\newcommand{\bes}{\begin{exs}}
\newcommand{\bb}{\begin{example}}
\newcommand{\bbs}{\begin{examples}}
\newcommand{\ba}{\begin{axiom}}
\newcommand{\bas}{\begin{assumption}}

\newcommand{\et}{\end{thm}}
\newcommand{\el}{\end{lem}}
\newcommand{\ec}{\end{cor}}
\newcommand{\ed}{\end{definition}}
\newcommand{\epp}{\end{prop}}
\newcommand{\er}{\end{rem}}
\newcommand{\en}{\end{note}}
\newcommand{\ee}{\end{ex}}
\newcommand{\ees}{\end{exs}}
\newcommand{\eb}{\end{example}}
\newcommand{\ebs}{\end{examples}}
\newcommand{\ea}{\end{axiom}}
\newcommand{\eas}{\end{assumption}}

\newcommand{\bp}{\begin{proof}}
\newcommand{\ep}{\end{proof}}
\newcommand{\eps}{\renewcommand{\qed}{}\end{proof}}

\newcommand{\bal}{\begin{align}}

\newcommand{\bi}[1][1.]{\begin{enumerate}[\upshape #1]}
\newcommand{\bia}[1][(1)]{\begin{enumerate}[\upshape #1]}
\newcommand{\bin}[1][1]{\begin{enumerate}[\upshape\bfseries #1]}
\newcommand{\bir}[1][(i)]{\begin{enumerate}[\upshape #1]}
\newcommand{\bic}[1][(i)]{\begin{enumerate}[\upshape\hspace{2\cma}#1]}
\newcommand{\bis}[2][1.]{\begin{enumerate}[\upshape\hspace{#2\parindent}#1]}
\newcommand{\ei}{\end{enumerate}}

\newcommand\ndots{\raise 0.47ex \hbox {,}\hskip0.06em\cdots %
     \raise 0.47ex \hbox {,}\hskip0.06em} 

\newcommand{\q}{\quad}
\newcommand{\qq}{\qquad}

\newcommand\nd{\noindent}

\newskip\Csmallskipamount                                                
\Csmallskipamount=\smallskipamount
\newskip\Cmedskipamount
\Cmedskipamount=\medskipamount
\newskip\Cbigskipamount
\Cbigskipamount=\bigskipamount

\newcommand\cvs{\vspace\Csmallskipamount}   
\newcommand\cvm{\vspace\Cmedskipamount}

\newskip\csa
\csa=\smallskipamount

\newskip\cma
\cma=\medskipamount

\newskip\cba
\cba=\bigskipamount

\newdimen\spt
\newcommand\citem{\cvs\advance\itemno by
1{(\romannumeral\the\itemno})\hskip3pt}
\newcommand{\bitem}{\cvm\nd\advance\itemno by
1{\bf\the\itemno}\hspace{\cma}}

\newcount\itemno
\newcommand{\lae}[1]{\label{E:#1}}
\newcommand{\lat}[1]{\label{T:#1}}

\newcommand{\laas}[1]{\label{Ass:#1}}

\newcommand{\rt}[1]{Theorem~\ref{T:#1}}

\newcommand{\ras}[1]{Assumption~\ref{Ass:#1}}

\newcommand{\re}[1]{\eqref{E:#1}}

\newcommand{\fre}[1]{\eqref{E:#1} on page~\tup{\pageref{E:#1}}}

\newcommand{\fras}[1]{Assumption~\ref{Ass:#1} on page~\tup{\pageref{Ass:#1}}}

\newskip\thmskip
\thmskip=\parindent

\newskip\hsk
\newenvironment{hinw}{\labelsep=0pt\begin{list}{}{\labelsep=0pt\itemindent=0pt\labelwidth=0pt\leftmargin=\parindent\rightmargin=0pt\partopsep=\cba}%
\item\it\nopagebreak\nopagebreak}%
{\end{list}}

\newcommand\bh{\begin{hinw}}
\newcommand{\eh}{\end{hinw}}

\newtheoremstyle{normal}
  {\cba}
  {\cba}
  {}
  {\thmskip}
  {\bfseries}
  {.}
  {\hsk}
  {}

\newtheoremstyle{abschnitt}
  {\cba}
  {\cba}
  {}
  {\thmskip}
  {\bfseries}
  {.}
  {\hsk}
  {}

\newtheoremstyle{italic}
  {\cba}
  {\cba}
  {\itshape}
  {\thmskip}
  {\bfseries}
  {.}
  {\hsk}
  {}

\newtheoremstyle{aufgaben}
  {\cba}
  {\cba}
  {}
  {}
  {\normalsize\bfseries}
  {.}
  {\hsk}
  {}

\newtheoremstyle{break}
  {\cba}
  {\cba}
  {\itshape}
  {}
  {\bfseries}
  {.}
  {\newline}
  {}

\swapnumbers
\theoremstyle{italic}
\newtheorem{thm}[subsection]{Theorem}
\newtheorem{lem}[subsection]{Lemma}
\newtheorem{prop}[subsection]{Proposition}
\newtheorem{cor}[subsection]{Corollary}

\theoremstyle{normal}
\newtheorem{rem}[subsection]{Remark}
\newtheorem{definition}[subsection]{Definition}
\newtheorem{example}[subsection]{Example}
\newtheorem{examples}[subsection]{Examples}
\newtheorem{ex}[subsection]{Exercise}
\newtheorem{note}[subsection]{}
\newtheorem{axiom}[subsection]{Axiom}
\newtheorem{assumption}[subsection]{Assumption}

\theoremstyle{aufgaben}
\newtheorem{exs}[subsection]{Exercises}

\swapnumbers

\numberwithin{equation}{section}
\numberwithin{figure}{section}

\newenvironment{textequation}[1][0.8]
{\begin{equation}
\begin{aligned}
\begin{minipage}{#1\linewidth}}
{\end{minipage}
\end{aligned}
\end{equation}
\ignorespacesafterend}

\newcommand{\btext}{\begin{textequation}}
\newcommand{\etext}{\end{textequation}}

\def\hinweis{\@startsection{subsection}{2}%
 \z@{0.7\linespacing\@plus 0.5\linespacing}{0.7\linespacing}%
{\normalfont\itshape\indent}}

\newcounter{hours}\newcounter{minutes}
\newcommand{\printtime}{%
\setcounter{hours}{\time/60}%
\setcounter{minutes}{\time-\value{hours}*60}%
\ifthenelse{\value{minutes}<10}{\thehours :0\theminutes}{\thehours:\theminutes}}

\usepackage[german,english]{babel}
\usepackage{graphicx}
\RequirePackage{amsmath}
\RequirePackage{bm}
\RequirePackage{amssymb}
\RequirePackage{upref}
\RequirePackage{amsthm}
\RequirePackage{enumerate}
\RequirePackage{pb-diagram}
\RequirePackage{amsfonts}
\RequirePackage[mathscr]{eucal}
\RequirePackage{verbatim}
\RequirePackage{xr}
\RequirePackage{graphicx}
\usepackage{calc}
\usepackage{xspace}
\usepackage{dsfont}
\usepackage{mathrsfs}

\makeatletter
\RequirePackage{color}
\newcommand{\ann}[1]{\renewcommand{\@makefnmark}{\mbox{$^{\color{red}{\@thefnmark}}$}}%
\footnote {#1}}
\makeatother

\RequirePackage{upref}
\RequirePackage{amsthm}
\RequirePackage{enumerate}
\usepackage[mathscr]{eucal}

\usepackage{xr-hyper}

\usepackage{calc}

\newlength{\oddsidemarginlength}
\newlength{\topmarginlength}

\newcounter{numberoflines}
\newcounter{tempcc}
\oddsidemargin=\oddsidemarginlength
\evensidemargin=\oddsidemargin
\topmargin=\topmarginlength
\usepackage[colorlinks=true,linkcolor=blue,citecolor=blue,urlcolor=blue]{hyperref}  

\begin{document}

\flushbottom


\title[A complete set of eigendistributions]{Deriving a complete set of eigendistributions for a gravitational wave equation describing the quantized interaction of gravity with a Yang-Mills field in case the Cauchy hypersurface is non-compact}

\author{Claus Gerhardt}
\address{Ruprecht-Karls-Universit\"at, Institut f\"ur Angewandte Mathematik,
Im Neuenheimer Feld 205, 69120 Heidelberg, Germany}
\email{\href{mailto:gerhardt@math.uni-heidelberg.de}{gerhardt@math.uni-heidelberg.de}}
\urladdr{\href{http://www.math.uni-heidelberg.de/studinfo/gerhardt/}{http://www.math.uni-heidelberg.de/studinfo/gerhardt/}}

%
\date{\today}
%


\begin{abstract} 
In a recent paper we quantized the interaction of gravity with a Yang-Mills and Higgs field and obtained as a result a gravitational wave equation in a globally hyperbolic spacetime. Assuming that the Cauchy hypersurfaces are compact we proved a spectral resolution for the wave equation by applying the method of separation of variables. In this paper we extend the results to the case when the Cauchy hypersurfaces are non-compact by considering a Gelfand triplet and applying the nuclear spectral theorem.
\end{abstract}

\maketitle

\tableofcontents

\setcounter{section}{0}
\section{Introduction}
In a recent paper \cite{cg:uf2} we quantized the interaction of gravity with a Yang-Mills and Higgs field and obtained as a result a gravitational wave equation of the form
\begin{equation}\lae{1.9}
\begin{aligned}
&\frac1{32}\frac {n^2}{n-1}\Ddot u-(n-1)t^{2-\frac4n}\D u-\frac n2t^{2-\frac4n}Ru+\al_1\frac n8 t^{2-\frac4n}F_{ij}F^{ij}u\\
&+\al_2 \frac n4 t^{2-\frac4n}\ga_{ab}\s^{ij}\F^a_i\F^b_iu+\al_2\frac n2 m t^{2-\frac4n}V(\F)u+nt^2\Lam u=0,
\end{aligned}
\end{equation}
in a globally hyperbolic spacetime
\begin{equation}
Q=(0,\un)\times \so
\end{equation}
describing the interaction of a given complete Riemannian metric $\s_{ij}$ in $\so$ with a  given Yang-Mills  and Higgs field; $R$ is the scalar curvature of $\s_{ij}$, $V$ is the potential of the Higgs field, $m$ a positive constant, $\al_1,\al_2$ are positive coupling constants and the other symbols should be self-evident. The existence of the time variable, and its range, is due to the quantization process.
\br
For the results and arguments in that paper it was completely irrelevant that the values of the Higgs field $\F$ lie in a Lie algebra, i.e., $\F$ could also be just an arbitrary scalar field, or we could consider a Higgs field as well as an another arbitrary scalar field. Hence, let us stipulate that the Higgs field could also be just an arbitrary scalar field. 
\er
If $\so$ is compact we also proved a spectral resolution of equation \re{1.9} by first considering a stationary version of the hyperbolic equation, namely, the elliptic eigenvalue equation
\begin{equation}\lae{1.11}
\begin{aligned}
&-(n-1)\D v-\frac n2Rv+\al_1\frac n8 F_{ij}F^{ij}v\\
&+\al_2 \frac n4 \ga_{ab}\s^{ij}\F^a_i\F^b_iv+\al_2\frac n2 m V(\F)v=\mu v.
\end{aligned}
\end{equation}
It has countably many solutions $(v_i,\mu_i)$ such that
\begin{equation}
\mu_0<\mu_1\le \mu_2\le \cdots,
\end{equation}
\begin{equation}
\lim \mu_i=\un.
\end{equation}
Let $v$ be an eigenfunction with eigenvalue $\mu>0$, then we  looked at solutions of \re{1.9} of the form
\begin{equation}\lae{1.6}
u(x,t)=w(t) v(x).
\end{equation}
$u$ is then a solution of \re{1.9} provided $w$ satisfies the implicit eigenvalue equation
\begin{equation}\lae{1.15}
-\frac1{32}\frac{n^2}{n-1}\Ddot w-\mu t^{2-\frac4n}w-nt^2\Lam w=0,
\end{equation}
where $\Lam$ is the eigenvalue.

This eigenvalue problem we also considered in a previous paper and proved that it has countably many solutions $(w_i,\Lam_i)$ with finite energy, i.e.,
\begin{equation}
\int_0^\un\{\abs{\dot w_i}^2+(1+t^2+\mu t^{2-\frac4n})\abs{w_i}^2\}<\un.
\end{equation}
More precisely, we proved, \cf \cite[Theorem 6.7]{cg:qgravity2},
\bt
Assume $n\ge 2$ and $\so$ to be compact and let $(v,\mu)$ be a solution of the eigenvalue problem \re{1.11} with $\mu>0$, then there exist countably many solutions $(w_i,\Lam_i)$ of the implicit eigenvalue problem \re{1.15} such that
\begin{equation}
\Lam_i<\Lam_{i+1}<\cdots <0,
\end{equation}
\begin{equation}
\lim_i\Lam_i=0,
\end{equation}
and such that the functions
\begin{equation}
u_i=w_i v
\end{equation}
are solutions of the wave equation \re{1.9}. The transformed eigenfunctions
\begin{equation}
\tilde w_i(t)=w_i(\lam_i^{\frac n{4(n-1)}}t), 
\end{equation}
where
\begin{equation}
\lam_i=(-\Lam_i)^{-\frac{n-1}n},
\end{equation}
form a basis of $L^2(\R[*]_+,\Cc)$ and also of the Hilbert space $H$ defined as the completion of $C^\un_c(\R[*]_+,\Cc)$ under the norm of the scalar product
\begin{equation}
\spd w{\tilde w}_1=\int_0^\un\{\bar w'\tilde w' +t^2\bar w\tilde w\},
\end{equation}
where a prime or a dot denotes differentiation with respect to $t$. 
\et
In this paper we want to extend this spectral resolution to the case when $\so$ is non-compact. Denote by $A$ the elliptic differential operator on the left-hand side of \re{1.11}, then, assuming that its coefficients are smooth with bounded $C^m$-norms for any $m\in\N$, we  have a self-adjoint operator in $H=L^2(\so)$ and a Gelfand triplet
\begin{equation}
\msc S\su H\su \msc S'
\end{equation}
such that we can apply the nuclear spectral theorem of Gelfand-Maurin leading to a complete set of eigendistributions 
\begin{equation}
f(\lam)\in\msc S',\qq\lam\in\Lam,
\end{equation}
 of $A$, where $\Lam$ is a measure space. For almost every $\lam\in\Lam$ we have $0\not=f(\lam)$ and $f(\lam)$ is a solution of the eigenvalue equation
 \begin{equation}
Af(\lam)=a(\lam)f(\lam)
\end{equation}
where
\begin{equation}
a:\Lam\ra \s(A)
\end{equation}
is a measurable function having $\s(A)$ as its essential range. Since the $f(\lam)$ are distributions  and A is uniformly elliptic and smooth, the $f(\lam)$ are also smooth, and since they are also tempered distributions we could prove that the eigenvalues satisfy
\begin{equation}
a(\lam)>0 
\end{equation}
for a.e.\ $\lam$. Hence, the separation of variables, described in \re{1.6}, can be applied with an eigenfunction $v$ be replaced by an eigendistribution $f$. Since all eigenvalues $a(\lam)$ are strictly positive this can be considered to be a spectral resolution of the wave equation. The smooth functions
\begin{equation}
u_i=w_if(\lam)
\end{equation}
are classical solutions of the wave equation \re{1.9} with bounded temporal energy and locally bounded spatial energy.

\section{The nuclear spectral theorem}
 We assume that $(\so,\s_{ij})$ is complete and that there exists a compact subset $K_0\su\so$ and a chart $(U_0,x)$ such that
\begin{equation}\lae{8.1}
\so\sminus K_0\su U_0
\end{equation}
and
\begin{equation}\lae{8.2}
\Om_0=x(U_0)=\R[n]\sminus \bar B_{R_0}(0).
\end{equation}
Moreover, we require
\bas\laas{8.1}
(i) The metric $\s_{ij}$ and the lower order coefficients of the elliptic operator on the left-hand side of equation of the equation \fre{1.11} are smooth with bounded $C^m$-norms for any $m\in\N$. We  call the elliptic operator $A$.

\cvm
(ii) The metric $\s_{ij}$ is uniformly elliptic.
\eas
The last assumption implies that the radial distance from a center $x_0\in K_0$,
\begin{equation}
r(x)=d(x,x_0),
\end{equation}
and the Euclidean distance $\abs x$ are equivalent in $\Om_0$, i.e., there are constants $c_1,c_2$ such that
\begin{equation}
r(x)\le c_1\abs x\le c_2r(x)\qq\A\, x\in\Om_0
\end{equation}
and hence the Schwartz space of rapidly decreasing test functions in $\so$ can be identified with the Schwartz space in $\R[n]$. We shall denote the Schwartz space by
\begin{equation}
\msc S=\msc S(S_0)
\end{equation}
and its dual space, the tempered distributions, by
\begin{equation}
\msc S'=\msc S'(\so).
\end{equation}
The topology of $\msc S$ is defined by a sequence of norms
\begin{equation}
\abs \f_{m,k}=\sup_{x\in\so}(1+r(x)^2)^k\sum_{\abs\al\le m}\abs{D^\al\f(x)}.
\end{equation}
$\msc S$ is a Fr\'echet space and also a nuclear space, \cf \cite[Example 5, p. 107]{schaefer:book}. The differential operator $A$ defined by the left-hand side of \fre{1.11} is a continuous map from $\msc S$ into $\msc S$ in view of \ras{8.1}. $A$ is a self-adjoint linear operator in $L^2(\so,\Cc)$ and
\begin{equation}
\msc S\su D(A)
\end{equation}
a dense subspace. By duality $A$ can also be defined on the dual space $\msc  S'$, namely, let $f\in\msc S'$, then
\begin{equation}
\spd{Af}\f=\spd f{A\f}\qq\A\,\f\in\msc S,
\end{equation}
where the self-adjointness of $A$ has been used.
\bd
$f\in\msc S'$ is said to be an eigendistribution of $A$ with eigenvalue $\m\in\R[]$, i.e.,
\begin{equation}
Af=\m f,
\end{equation}
iff
\begin{equation}
\spd{Af}\f=\mu\spd f\f\qq\A\,\f\in\msc S,
\end{equation}
or equivalently, iff
\begin{equation}
\spd{f}{A\f}=\mu\spd f\f\qq\A\,\f\in\msc S.
\end{equation}
\ed
A setting where we have a self-adjoint operator $A$ in a separable Hilbert space $H$, a dense subspace
\begin{equation}\lae{8.13}
E\su H
\end{equation}
which is also a nuclear space (in a finer topology) with dual space $E'$ such that
\begin{equation}
E\su H\su E',
\end{equation}
where the imbedding of $E$ into $H$ is continuous, a property which we already specified by speaking of a \tit{finer} topology, and where, moreover,
\begin{equation}\lae{8.15}
A:E\ra E
\end{equation}
is continuous, is known as a \tit{rigged Hilbert space} setting, though, usually, $E'$ is replaced by the space of antilinear functionals. However, since we do not use Dirac's ket notation, we shall consider $E'$.

In such a framework a nuclear spectral theorem has been proved by Gelfand and Maurin, \cf \cite[Theorem 5', p. 126]{gelfand:book}, \cite[Satz 2]{maurin:rigged} and \cite[Chap. XVIII, p.\ 333]{maurin:book} which we shall formulate and prove for a single self-adjoint operator $A$ and not for a family of strongly commuting operators. The proof closely follows the one given by Maurin in \cite[Satz 2]{maurin:rigged}. Since this paper is written in German we like to include a proof for the convenience of the reader. 
\bt[Maurin]
Let $H$ be a separable complex Hilbert space, $A$ a densely defined self-adjoint operator,  $E\su H$ a dense subspace which also carries topology such that it is a nuclear space and assume that the imbedding in \re{8.13} and the map $A$ in \re{8.15} are continuous, then there exists a locally compact measure space  $\Lam$, a finite positive measure $\mu$,  a measurable function 
\begin{equation}
a:\Lam\ra \s(A)\su\R[],
\end{equation}
and a unitary operator
\begin{equation}
U:H\ra L^2(\Lam,\Cc,\mu)
\end{equation}
such that, if we set
\begin{equation}
\hat u=Uu,\qq\A\, u\in H,
\end{equation}
\begin{equation}
\hat A=UAU^{-1},
\end{equation}
we have
\begin{equation}
u\in D(A)\eqv a\hat u\in L^2(\Lam, \mu),
\end{equation}
\begin{equation}\lae{8.21}
\hat A\hat u=a\hat u\qq\A\, u\in D(A)
\end{equation}
and for $\mu$ a.e.\ $\lam\in\Lam$ the mapping
\begin{equation}\lae{8.22}
f(\lam):\f\in E\ra \hat\f(\lam)\in\Cc
\end{equation}
is continuous in $E$ and does not vanish identically, i.e.,
\begin{equation}
0\not=f(\lam)\in E'
\end{equation}
and hence
\begin{equation}
\hat\f(\lam)=\spd{f(\lam)}\f\qq\A\f\in E.
\end{equation}
Moreover,  \re{8.21} implies
\begin{equation}
\begin{aligned}
\spd{f(\lam)}{A\f}=\hat A\hat\f(\lam)=a(\lam)\hat\f(\lam)=a(\lam)\spd{f(\lam)}\f\q\A\, \f\in E
\end{aligned}
\end{equation}
for a.e. $\lam\in\Lam$, or equivalently,
\begin{equation}
Af(\lam)=a(\lam)f(\lam)\qq \tup{for a.e. } \lam\in \Lam.
\end{equation}
The generalized eigenvectors $f(\lam)$ are complete, since
\begin{equation}
\norm\f^2=\norm{\hat\f}^2=\int_\Lam\abs{\hat\f(\lam)}^2d\mu\qq\A\,\f\in E,
\end{equation}
and hence,
\begin{equation}
\hat\f(\lam)=0\qq\tup{for a.e. }\lam\in \Lam,
\end{equation}
is equivalent to $\f=0$.
\et
\bp
The first part of the theorem is due to the multiplicative form of the spectral theorem, \cf \cite[Theorem VIII.4, p.\ 260]{reed-simon}. Let us remark that we used a different version of von Neumann's spectral theorem than Maurin which simplifies the proof slightly, especially the completeness part. Note that the spectrum
\begin{equation}
\s(A)=\s(\hat A)
\end{equation}
is the essential range of $a$.

To prove \re{8.22} and the following claims, we observe that the imbedding
\begin{equation}\lae{8.30}
j:E\ra H
\end{equation}
is continuous and therefore also nuclear, hence there is a semi-norm $\norm\cdot_p$ on $E$ sequences $u_k\in H$, $f_k\in E'$ such that
\begin{equation}
j(\f)=\sum_k\spd{f_k}\f u_k\qq\A\,\f\in E
\end{equation}
and
\begin{equation}\lae{8.32}
\sum_k\norm{f_k}_{-p}\norm {u_k}=\sum_k\norm{f_k}_{-p}\norm{\hat u_k}<\un,
\end{equation}
where $\norm\cdot_{-p}$ is the dual norm in $E'$
\begin{equation}
\norm{f_k}_{-p}=\sup_{\norm\f_p=1}\abs{\spd{f_k}\f}.
\end{equation}
We shall show that the mapping in \re{8.30}, which, when composed with $U$, can now be expressed as
\begin{equation}\lae{8.34}
\f\ra \hat\f(\lam)=\sum_k\spd{f_k}\f \hat u_k(\lam)
\end{equation}
is continuous in $E$ and not identically $0$ for a.e. $\lam\in\Lam$.

Indeed, without loss of generality we may assume  
\begin{equation}
\norm{u_k}=1
\end{equation}
to deduce from \re{8.32} 
\begin{equation}\lae{8.36}
\begin{aligned}
\sum_k\norm{f_k}_{-p}&=\sum_k\norm{f_k}_{-p}\norm{u_k}=\sum_k\norm{f_k}_{-p}\norm{u_k}^2\\
&=\sum_k\norm{f_k}_{-p}\int_\Lam \abs{\hat u_k(\lam)}^2=\int_\Lam\sum_k\norm{f_k}_{-p}\abs{\hat u_k(\lam)}^2<\un,
\end{aligned}
\end{equation}
hence
\begin{equation}\lae{8.37}
\sum_k\norm{f_k}_{-p}\abs{\hat u_k(\lam)}^2\equiv c_1(\lam)^2<\un
\end{equation}
for a.e. $\lam\in \Lam$, and
\begin{equation}
c_1(\cdot)\in L^2(\Lam,\mu).
\end{equation}

To prove \re{8.22} we now estimate
\begin{equation}
\begin{aligned}
\abs{\hat\f(\lam)}^2&=\absb{\sum_k\spd{f_k}\f\hat u_k(\lam)}^2\le \big(\sum_k\norm{f_k}_{-p}\norm\f_p\abs{\hat u_k(\lam)}\big)^2\\
&=\big(\sum_k\norm{f_k}_{-p}^\frac12\norm\f_p\norm{f_k}_{-p}^\frac12\abs{\hat u_k(\lam)}\big)^2\\
&\le \big(\sum_k\norm{f_k}_{-p}\norm\f_p^2\big)\big(\sum_k\norm{f_k}_{-p}\abs{\hat u_k(\lam)}^2\big)\\
&=c_1(\lam)^2\sum_k\norm{f_k}_{-p}\norm\f_p^2<\un,
\end{aligned}
\end{equation}
in view of \re{8.36} and \re{8.37}. 

The fact that the mapping in \re{8.34} does not vanish identically for a.e. $\lam\in\Lam$ is proved in the lemma below. This completes the proof of the theorem, since the  other properties are evident.
\ep
\bl
The mapping \re{8.34} does not vanish identically in $E$ for a.e. $\lam\in\Lam$.
\el
\bp
We argue by contradiction and assume that there exists a measurable set $\Lam_0\su\Lam$ with positive measure such that
\begin{equation}
\hat \f(\lam)=0\qq\A\, (\lam,\f)\in \Lam_0\times E.
\end{equation}
Let $\chi_0$ be the characteristic function of $\Lam_0$ and set
\begin{equation}
u_0=U^{-1}\chi_0,
\end{equation}
then
\begin{equation}
0\not=u_0\in H.
\end{equation}
Let $\f_k\in E$ be sequence converging to $u_0$, then
\begin{equation}
\norm{u_0}^2=\lim_k\spd{\f_k}{u_0}=\lim_k\int_{\Lam}\bar{\hat\f}_k\chi_0=0,
\end{equation}
a contradiction.
\ep
\section{The eigendistributions are smooth functions}

In our case $E=\msc S$ and $A$ is a uniformly elliptic linear differential operator with smooth coefficients. Hence, we can prove:
\bt
Let $A$ satisfy the \fras{8.1}, then the solutions $f(\lam)\in \msc S'$ of the eigenvalue problem
\begin{equation}\lae{9.1}
Af(\lam)=\mu f(\lam)
\end{equation}
belong to $C^\un(\so)$ and for each $m\in\N$ and $R>0$ $f(\lam)$ can be estimated by
\begin{equation}\lae{9.2}
\abs{f(\lam)}_{m,B_R(x_0)}\le c_m R^N\norm{f(\lam)}_{-p},
\end{equation}
where $\norm\cdot_p$ is one of the defining norms in $\msc S$ such that
\begin{equation}
\norm{f(\lam)}_{-p}=\sup_{\norm\f_p=1}\abs{\spd{f(\lam)}\f}
\end{equation}
and $N$ depends on $n$, $\norm\cdot_p$, $A$ and $\so$, while $c_m$  depends on $m$, $A$ the eigenvalue $\mu$ and on $\so$. $B_R(x_0)$ is a geodesic ball of radius $R$ for a fixed $x_0\in K_0\su\so$.
\et
\bp
First we note that we can absorb the right-hand side of the eigenvalue equation into the left-hand side and simply consider the equation
\begin{equation}\lae{9.4}
Af(\lam)=0.
\end{equation}
Hence, it is well-known that the distributional solutions is smooth and equation \re{9.4} can be understood in the classical sense, see e.g., \cite[Theorem 3.2, p.125]{lions:book}.

The important estimate \re{9.2} is due to the fact that $f(\lam)$ is a tempered distribution. Since $f(\lam)\in\msc S'$ we have
\begin{equation}\lae{9.5}
\abs{\spd{f(\lam)}\f}\le c \sup_{x\in\so}(1+r(x)^2)^k\sum_{\abs\al\le m_0}\abs{D^\al\f(x)}\equiv c\norm\f_p
\end{equation}
and the dual norm
\begin{equation}
\norm{f(\lam)}_{-p}=c.
\end{equation}
To prove \re{9.2} we fix $m\in\N$ and assume that
\begin{equation}
\abs{f(\lam)}_{m,B_{R_1}(x_0)}\le c_0,
\end{equation}
for some sufficiently large radius $R_1$ such that we only have to prove the estimate in the domain
\begin{equation}
B_R(0)\sminus \bar B_{R_0}(0),
\end{equation}
where we now consider Euclidean balls, \cf the assumptions in \re{8.1} and \fre{8.2}. Hence we may consider equation \re{9.4} to be a uniformly elliptic equation in an exterior region of Euclidean space with smooth coefficients.

Let $R>R_0$, then we first prove a priori estimates for $f(\lam)$ in smalls balls
\begin{equation}
B_\rho(y)\Su B_{2R}(0)\sminus B_{R_0}(0),
\end{equation}
where
\begin{equation}
2\rho<\rho_0\le 1
\end{equation}
and $\rho_0$ is fixed.

Let
\begin{equation}
H^{m,2}_0(\Om),\qq m\in\N,
\end{equation}
be the usual Sobolev spaces, where
\begin{equation}
\Om\su\R[n]
\end{equation}
is an open set, to be defined as the completion of $C^\un_c(\Om)$ under the norm
\begin{equation}
\norm\f_{m,2}^2=\int_\Om\sum_{\abs\al\le m}\abs{D^\al\f}^2.
\end{equation}
$H^{m,2}_0(\Om)$ is a Hilbert space. Its dual space is denoted by
\begin{equation}
H^{-m,2}(\Om)
\end{equation}
and its elements are the distributions $f\in \msc D'(\Om)$ which can be written in the form
\begin{equation}
f=\sum_{\abs{\al}\le m}D^\al u_\al,
\end{equation}
where
\begin{equation}
u_\al\in L^2(\Om)
\end{equation}
and the dual norm of $f$ is equal to
\begin{equation}
\norm f_{-m,2}=\big(\sum_{\abs\al\le m}\norm{u_\al}_2^2\big)^\frac12.
\end{equation}
The Sobolev imbedding theorem states that
\begin{equation}
m>\frac n2\q\im\q  H^{m,2}_0(\Om)\hra C^0(\Om)
\end{equation}
such that
\begin{equation}
\abs u_0\le c \norm u_{m,2}\qq\A\, u\in H^{m,2}_0(\Om),
\end{equation}
where $c$ only depends on $m$ and $n$.

As a corollary we deduce
\begin{equation}
m>\frac n2\q\im\q  H^{m+m_0,2}_0(\Om)\hra C^{m_0,0}(\Om)
\end{equation}
with a corresponding estimate
\begin{equation}
\abs u_{m_0,0}\le c \norm u_{m+m_0,2},
\end{equation}
where $c=c(n,m,m_0)$.

Hence, for any ball
\begin{equation}
B_{\rho_0}(y)\su B_{2R}(0)
\end{equation}
$f(\lam)$ can be considered to belong to
\begin{equation}
f(\lam)\in H^{-(m_0+n),2}(B_{\rho_0}(y))
\end{equation}
with norm
\begin{equation}
\norm{f(\lam)}_{-(n+m_0),2}\le c R^{2k}
\end{equation}
in view of the estimate \re{9.5}, where we also assume $R_0>1$; the constant $c$ depends on $n$, $m_0$, $k$ and the constant in \re{9.5}.

From the proofs of \cite[Theorem 3.1, p. 123]{lions:book} and \cite[Theorem 3.2, p. 125]{lions:book} we then deduce that for any $m\in\N$ there exists $\rho<\rho_0$, $\rho$ depending only on the Lipschitz constant of the metric $\s_{ij}$, $m,n$ and $m_0$ such that the $C^m$-norm of the solution $f(\lam)$ of equation \re{9.4} can be estimated by
\begin{equation}
\abs{f(\lam)}_{m,B_{\rho}(y)}\le c_\rho R^{2k},
\end{equation}
where $c_\rho$ also depends on the $C^m$-norms of the coefficients of $A$ and on the ellipticity constants.

Now
\begin{equation}
(4R)^n2^n\rho^{-n}
\end{equation}
balls
\begin{equation}
B_\rho(y)\su B_{2R}(0)
\end{equation}
cover the closed ball $\bar B_R(0)$, hence we conclude
\begin{equation}
\abs{f(\lam)}_{m,B_R(0)\sminus K_0}\le c R^{2k+n},
\end{equation}
where $c=c(\rho,m,m_0,n,A)$.
\ep
\section{The positivity of the eigenvalues}
To apply the separation of variables method to find a complete set of eigensolutions for the wave equation the eigenvalues of the elliptic operator have to be positive. In this section we shall prove that eigenvalues of the eigenvalue equation \fre{9.1} are always strictly positive provided some rather weak assumptions are satisfied.

Let us start with the following lemma:
\bl
Let $A$ be the differential operator on the left-hand side of \fre{1.11} and let us write the operator in the form
\begin{equation}
Av=-(n-1)\D v -\frac n2 Rv +Gv+\al_2\frac n2 m V(\F)v,
\end{equation}
where
\begin{equation}
0\le G=\al_1\frac n8 F_{ij}F^{ij}+\al_2\frac n2 \ga_{ab}\s^{ij} \F^a_i\F^b_j.
\end{equation}
Assume there are positive constants $\e_0$, $\de$, $m_0$ and $R_1$ such that
\begin{equation}\lae{10.3}
-\frac n2 R+G+\al_2\frac n2 m_0 V\ge \e_0 r^{-2+\de}\qq\A\, x\notin B_{R_1}(x_0),
\end{equation}
where $x_0\in K_0$ is fixed and $r$ is the geodesic distance to $x_0$, then there exists $m_1\ge m_0$ such that for all $m\ge m_1$ the quadratic form of $A$ satisfies
\begin{equation}
\int_{B_{R_1}(x_0)}\norm u^2\le\spd{Au}u\qq\A\, u\in H^{1,2}(\so),
\end{equation}
provided
\begin{equation}\lae{10.5}
V(\F)>0 \q \tup{a.e.\ in} \;\so.
\end{equation}
\el
\bp
The ball $B_{R_1}(x_0)$ is bounded, hence the imbedding of
\begin{equation}
H^{1,2}(B_{R_1}(x_0))\hra L^2(B_{R_1}(x_0))
\end{equation}
is compact and we can apply a compactness lemma to conclude that for any $\e>0$ there is a constant $c_\e$ such that
\begin{equation}
\int_{B_{R_1}(x_0)}\abs u^2\le \e \int_{B_{R_1}(x_0)}\abs{Du}^2+c_\e\int_{B_{R_1}(x_0)}V(\F)\abs u^2
\end{equation}
for all $u\in H^{1,2}(B_{R_1}(x_0))$, in view of the assumption \re{10.5}, \cf \cite[Lemma 7.5]{cg:uf2}. 

Hence, we deduce
\begin{equation}
\begin{aligned}
\int_{B_{R_1}(x_0)}\abs u^2&\le(n-1)\int_{B_{R_1}(x_0)}\abs{Du}^2+\int_{B_{R_1}(x_0)}(-\frac n2 R+G)\abs u^2\\
&\q +\al_2\frac n2 m \int_{B_{R_1}(x_0)}V(\F)\abs u^2
\end{aligned}
\end{equation}
for all $u\in H^{1,2}(B_{R_1}(x_0))$ provided $m$ is sufficiently large
\begin{equation}
m\ge m_1.
\end{equation}
Choosing $m_1\ge m_0$ completes the proof of the lemma because of the assumption \re{10.3}.
\ep
\bt\lat{10.2}
Under the assumptions of the preceding lemma and the general provisions in \re{8.1}, \re{8.2} and \fras{8.1} the eigenvalue equation \fre{9.1} is only solvable if $\mu>0$.
\et
\bp
Since the quadratic form of $A$ is positive we immediately infer
\begin{equation}
\s(A)\su \R[]_+,
\end{equation}
hence the eigenvalue $\mu$ in \re{9.1} has to satisfy
\begin{equation}
0\le\mu
\end{equation}
so that we have to exclude the case 
\begin{equation}
\mu=0.
\end{equation}
We argue by contradiction. Let
\begin{equation}
f\in \msc S'\ii C^\un (\so)
\end{equation}
be a solution of
\begin{equation}\lae{10.14}
Af=0,
\end{equation}
then we shall prove
\begin{equation}
f=0.
\end{equation}
Let $k\in\N$ and $R>R_1$ be large and let $\h$ be defined by
\begin{equation}
\h(x)=\begin{cases}
R^{-k},&\abs x\le R,\\
\abs x^{-k}, &\abs x>R,
\end{cases}
\end{equation}
Then
\begin{equation}
f\h^2\in H^{1,2}(\so),
\end{equation}
in view of the estimate \fre{9.2}, which can be rephrased to
\begin{equation}
\sum_{\abs\al\le m}\abs{D^\al f(x)}\le c_m  \abs x^N\qq\A\, \abs x>R_0.
\end{equation}
Here, we use the Euclidean distance.

Multiplying \re{10.14} by $f\h^2$ and integrating by parts yields
\begin{equation}
\begin{aligned}
0&\ge \int_\so\{(n-1) \abs{Df}^2\h^2-\frac n2 R\abs f^2+G\abs f^2+\al_2\frac n2 m \abs f^2\}\h^2\\
&\q -(n-1)\int_{\so\sminus B_{R_1}(x_0)}2\abs {Df}\abs f\h\abs{D\h}\\
&\ge \int_{B_R}\{(n-1) \abs{Df}^2-\frac n2 R\abs f^2+G\abs f^2+\al_2\frac n2 m \abs f^2\}R^{-2k}\\
&\q+ \int_{\so\sminus B_R}\{\e_0\abs x^{-2+\de}-c\abs x ^{-2}\}\abs f^2\h^2,
\end{aligned}
\end{equation}
where $c$ is a fixed constant depending only on the metric $\s_{ij}$ and $k$.

The first integral is strictly positive unless $f$ vanishes in $B_R(x_0)$, and the difference in the braces is also strictly positive if $R$ is large enough. Hence we conclude
\begin{equation}
f\equiv 0.
\end{equation}
\ep

We can now prove a spectral resolution of the hyperbolic equation \fre{1.9} by choosing an eigendistribution $f=f(\lam)$ with  eigenvalue $\mu=a(\lam)$ and look at solutions of \re{1.9} of the form
\begin{equation}
u(x,t)=w(t) f(x).
\end{equation}
$u$ is then a solution of \re{1.9} provided $w$ satisfies the implicit eigenvalue equation
\begin{equation}\lae{7.9}
-\frac1{32}\frac{n^2}{n-1}\Ddot w-\mu t^{2-\frac4n}w-nt^2\Lam w=0,
\end{equation}
where $\Lam$ is the eigenvalue.

This eigenvalue problem we also considered in a previous paper and proved that it has countably many solutions $(w_i,\Lam_i)$ with finite energy, i.e.,
\begin{equation}
\int_0^\un\{\abs{\dot w_i}^2+(1+t^2+\mu t^{2-\frac4n})\abs{w_i}^2\}<\un,
\end{equation}
\cf \cite[Theorem 6.7]{cg:qgravity2}.

We can then extend the spectral resolution which we proved in \cite[Theorem 1.7]{cg:uf2} for a compact Cauchy hypersurface $\so$ to the case when $\so$ is non-compact:
\bt\lat{7.2}
Assume $n\ge 2$ and  let $\so$ and the elliptic differential operator $A$ satisfy the assumptions of the \rt{10.2}. Pick any  solution $(f,\mu)$ of the eigenvalue problem \re{9.1}, then there exist countably many solutions $(w_i,\Lam_i)$ of the implicit eigenvalue problem \re{7.9} such that
\begin{equation}
\Lam_i<\Lam_{i+1}<\cdots <0,
\end{equation}
\begin{equation}
\lim_i\Lam_i=0,
\end{equation}
and such that the functions
\begin{equation}
u_i=w_i f
\end{equation}
are solutions of the wave equations \fre{1.9}. The transformed eigenfunctions
\begin{equation}
\tilde w_i(t)=w_i(\lam_i^{\frac n{4(n-1)}}t), 
\end{equation}
where
\begin{equation}
\lam_i=(-\Lam_i)^{-\frac{n-1}n},
\end{equation}
form a basis of $L^2(\R[*]_+,\Cc)$ and also of the Hilbert space $H$ defined as the completion of $C^\un_c(\R[*]_+,\Cc)$ under the norm of the scalar product
\begin{equation}
\spd w{\tilde w}_1=\int_0^\un\{\bar w'\tilde w' +t^2\bar w\tilde w\},
\end{equation}
where a prime or a dot denotes differentiation with respect to $t$. 
\et

\br
This result is the best we can achieve under the present assumptions. In order to prove a mass gap, i.e., prove an  estimate of the form
\begin{equation}
0<\e_0\le \mu
\end{equation}
for ally eigenvalues $\mu$ of the eigenvalue equation \fre{9.1} we have to strengthen our assumptions on the zero order terms: Instead of the assumption \re{10.3} we have to require
\begin{equation}\lae{10.31}
-\frac n2 R+G+\al_2\frac n2 m_0 V\ge \e_0>0\qq\A\, x\notin B_{R_1}(x_0),
\end{equation}
then we immediately would derive a mass gap

An even stronger estimate of the form
\begin{equation}\lae{10.32}
-\frac n2 R+G+\al_2\frac n2 m_0 V\ge \e_0 r^{\de}\qq\A\, x\notin B_{R_1}(x_0),
\end{equation}
with $\de>0$, would yield that the operator $A$ would have a pure point spectrum since the quadratic form
\begin{equation}
\spd{Au}u
\end{equation}
would then be compact relative to the $L^2$-scalar product and we would be in the same situation as if $\so$ would be compact.
\er

\bibliographystyle{hamsplain}
\providecommand{\bysame}{\leavevmode\hbox to3em{\hrulefill}\thinspace}
\providecommand{\href}[2]{#2}



\end{document}